\RequirePackage{amsmath}
\documentclass[graybox]{svmult}

\usepackage{amssymb}       
\usepackage{helvet}         
\usepackage{courier}        
\usepackage{type1cm}        

\usepackage{makeidx}         
\usepackage{graphicx}        
\usepackage{multicol}        
\usepackage[bottom]{footmisc}

\usepackage{psfrag,bbm,epsf}
\usepackage{xspace,subfigure}
\usepackage{comment}
\usepackage{amssymb, xcolor,enumerate,float,mathtools}
\usepackage{algpseudocode}
\usepackage{algorithm, algorithmicx}
\usepackage{enumerate,multirow,booktabs}
\usepackage{url}
\usepackage{amsrefs}

\algnewcommand\algorithmicparam{\textbf{Parameters:}}
\algnewcommand\PARAM{\item[\algorithmicparam]}
\algnewcommand\algorithmicinput{\textbf{Input:}}
\algnewcommand\INPUT{\item[\algorithmicinput]}
\algnewcommand\algorithmicoutput{\textbf{Output:}}
\algnewcommand\OUTPUT{\item[\algorithmicoutput]}
\algnewcommand\RETURN{\State \textbf{Return }}

\newcommand{\reals}{\mathbb{R}}

\newcommand{\va}{\boldsymbol{a}}
\newcommand{\vb}{\boldsymbol{b}}
\newcommand{\vc}{\boldsymbol{c}}
\newcommand{\vx}{\boldsymbol{x}}
\newcommand{\vX}{\boldsymbol{X}}
\newcommand{\vt}{\boldsymbol{t}}

\newcommand{\vu}{\boldsymbol{u}}
\newcommand{\vv}{\boldsymbol{v}}

\newcommand{\fraku}{\mathfrak{u}}
\newcommand{\frakv}{\mathfrak{v}}

\newcommand{\vinfty}{\boldsymbol{\infty}}
\newcommand{\vzero}{\boldsymbol{0}}
\newcommand{\vPsi}{\boldsymbol{\Psi}}
\newcommand{\dif}{{\rm d}}
\DeclareMathOperator{\sign}{sign}
\DeclareMathOperator*{\argmax}{argmax}
\newcommand{\Udes}{\mathcal{U}}
\newcommand{\Xdes}{\mathcal{X}}

\newcommand{\cm}{\mathcal{M}}
\newcommand{\ch}{\mathcal{H}}

\newcommand{\Ftar}{F}
\newcommand{\ftar}{\varrho}
\newcommand{\cube}{\ensuremath{(0,1)^d}}
\newcommand{\bbone}{\mathbbm{1}}
\newcommand{\Ex}{\mathbb{E}}
\newcommand{\ip}[3][{}]{\ensuremath{\left \langle #2, #3 \right \rangle_{#1}}}
\newcommand{\norm}[2][{}]{\ensuremath{\left \lVert #2 \right \rVert}_{#1}}
\DeclareMathOperator{\cov}{cov}
\DeclareMathOperator{\err}{err}
\newcommand{\unif}{\textup{unif}}
\newcommand{\normal}{\textup{normal}}
\newcommand{\frakd}{\mathfrak{d}}
\newcommand{\tK}{\widetilde{K}}
\newcommand{\tOmega}{\widetilde{\Omega}}
\newcommand{\tvarrho}{\widetilde{\varrho}}
\newcommand{\Order}{\mathcal{O}}
\newcommand{\TOL}{\text{TOL}}
\newcommand{\vT}{\boldsymbol{T}}
\newcommand{\vone}{\boldsymbol{1}}
\newcommand{\vgamma}{\boldsymbol{\gamma}}
\newcommand{\figdir}{} 

\def\abs#1{\ensuremath{\left \lvert #1 \right \rvert}}

\usepackage{bm} 

\makeindex             

\begin{document}

\title*{Is a Transformed Low Discrepancy Design Also Low Discrepancy?}
\author{Yiou Li, Lulu Kang, and Fred J. Hickernell}
\institute{Yiou Li \at DePaul University, 2320 N. Kenmore Avenue, Chicago, IL 60614 \email{yli139@depaul.edu}
\and Lulu Kang \at Illinois Institute of Technology, RE 208, 10 W. 32nd Street, Chicago, IL 60616 \email{lkang2@iit.edu} 
\and 
Fred J. Hickernell \at Illinois Institute of Technology, RE 208, 10 W. 32nd Street, Chicago, IL 60616 \email{hickernell@iit.edu}}
%
%
\maketitle

\abstract{Experimental designs intended to match arbitrary target distributions are typically constructed via a variable transformation of a uniform experimental design.
The inverse distribution function is one such transformation.
The discrepancy is a measure of how well the empirical distribution of any design matches its target distribution.
This chapter addresses the question of whether a variable transformation of a low discrepancy uniform design yields a low discrepancy design for the desired target distribution.
The answer depends on the two kernel functions used to define the respective discrepancies.  
If these kernels satisfy certain conditions, then the answer is yes.  
However, these conditions may be undesirable for practical reasons.  
In such a case, the transformation of a low discrepancy uniform design may yield a design with a large discrepancy. 
We illustrate how this may occur.
We also suggest some remedies. 
One remedy is to ensure that the original uniform design has optimal one-dimensional projection, but this remedy works best if the design is dense, or in other words, the ratio of sample size divided by the dimension of the random variable is relatively large. 
Another remedy is to use the transformed design as the input to a coordinate-exchange algorithm that optimizes the desired discrepancy, and this works for both dense or sparse designs. 
The effectiveness of these two remedies is illustrated via simulation.}

\section{Introduction}

Professor Kai-Tai Fang and his collaborators have demonstrated the effectiveness of low discrepancy points as space filling designs \cites{FangHic07a, FangEtal19a, FanLiSud06, FanWan94}. 
They have promoted discrepancy as a quality measure for statistical experimental designs to the statistics, science, and engineering communities \cites{FanMa01b, FanMaWin02, FanMuk00, FanMa01a}. 

Low discrepancy uniform designs, $\Udes = \{\vu_i\}_{i=1}^N$, are typically constructed so that their empirical distributions, $F_\Udes$, approximate $F_\unif$, the uniform distribution on the unit cube, \cube. 
The discrepancy measures the magnitude of $F_\unif-F_\Udes$. 
The uniform design is a commonly used space filling design for computer experiments \cite{FanLiSud06} and can be constructed using  JMP\textsuperscript{\textregistered} \cite{sall2012jmp}.

When the target probability distribution for the design, $\Ftar$, defined over the experimental domain $\Omega$, is \emph{not} the uniform distribution on the unit cube, then the desired design, $\Xdes$, is typically constructed by transforming a low discrepancy uniform design, i.e., 
\begin{equation} \label{eq:transPts}
\Xdes = \{\vx_i\}_{i=1}^N = \{\vPsi(\vu_i)\}_{i=1}^N = \vPsi(\Udes), \qquad \vPsi: \cube \to \Omega.
\end{equation}
Note that $\Ftar$ may differ from $F_\unif$ because $\Omega \ne \cube$ and/or $\Ftar$ is non-uniform.  
A natural transformation, $\vPsi(\vu)=\bigl(\Psi_1(u_1),\ldots,\Psi_d(u_d) \bigr)$, when $\Ftar$ has independent marginals, is the inverse distribution transformation:
\begin{equation}\label{eq:inverse}
\Psi_j(u_j) = F_j^{-1}(u_j), \quad j =1, \ldots, d, \qquad \text{where } \Ftar(\vx) = F_1(x_1) \cdots F_d(x_d).
\end{equation}
A number of transformation methods for different distributions can be found in \cite{DEVROYE200683} and \cite{FanWan94}*{Chapter 1}.

This chapter addresses the question of whether the design $\Xdes$ resulting from transformation \eqref{eq:transPts} of a low discrepancy design, $\Udes$, is itself low discrepancy with respect to the target distribution $\Ftar$. 
In other words, 
\begin{equation} \label{eq:BigQ} \tag{Q}
\text{\emph{does small $F_\unif - F_\Udes$ imply small $\Ftar - F_\Xdes$?}}
\end{equation}
We show that the answer may be yes or no, depending on how the question is understood.  
We discuss both cases.  
For illustrative purposes, we consider the situation where $\Ftar$ is the standard multivariate normal distribution, $F_\normal$.

In the next section, we define the discrepancy and motivate it from three perspectives.  In Section \ref{sec:WhenYes} we give a simple condition under which the answer to \eqref{eq:BigQ} is yes.  But, in Section \ref{sec:DoTransformedPts} we show that under more practical assumptions the answer to \eqref{eq:BigQ} is no.  An example illustrates what can go wrong.  Section \ref{sec:CoordEx} provides a coordinate exchange algorithm that improves the discrepancy of a candidate design.  Simulation results illustrate the performance of this algorithm.  We conclude with a brief discussion.

\section{The Discrepancy} \label{sec:discrepancy}

Experimental design theory based on  discrepancy assumes an experimental region, $\Omega$, and a target probability distribution, $\Ftar:\Omega \to [0,1]$, which is known a priori. We assume that $\Ftar$ has a probability density, $\ftar$.  It is convenient to also work with measures, $\nu$, defined on $\Omega$.  If $\nu$ is a probability measure, then the associated probability distribution is given by $F(\vx) = \nu((-\vinfty,\vx])$.  The Dirac measure, $\delta_{\vx}$ assigns unit measure to the set $\{\vx\}$ and zero measure to sets not containing $\vx$.  A design, $\Xdes = \{\vx_i\}_{i=1}^N$, is a finite set of points with empirical distribution $F_{\Xdes}  = N^{-1} \sum_{i=1}^N \bbone_{(-\vinfty,\vx_i]}$ and empirical measure $\nu_{\Xdes}  = N^{-1} \sum_{i=1}^N \delta_{\vx_i}$.

Our notation for discrepancy takes the form of 
\[
 D(F_{\Xdes},F,K), \ D(\Xdes,F,K), \ D(\Xdes,\varrho,K), \ D(\Xdes,\nu,K), \ D(\nu_{\Xdes},\nu,K), \text{ etc.}, 
\]
all of which mean the same thing.  The first argument always refers to the design, the second argument always refers to the target, and the third argument is a symmetric, positive definite kernel, which is explained below.  We abuse the discrepancy notation because sometimes it is convenient to refer to the design as a set, $\Xdes$, other times by its empirical distribution, $F_{\Xdes}$, and other times by its empirical measure, $\nu_{\Xdes}$.  Likewise, sometimes it is convenient to refer the target as a probability measure,  $\nu$, other times by its distribution function, $F$, and other times by its density function, $\varrho$.

\begin{table}
\centering
\caption{Three interpretations of the discrepancy.}
\vspace{-3ex}
\begin{equation*}
\begin{array}{c@{\qquad}ccc}
\text{Kernel Interpretation} & \text{Discrepancy } D(\Xdes,\nu,K) = D(\Xdes,\varrho,K)\\
\toprule \\[-1ex]
K(\vt,\vx) = \ip[\cm]{\delta_{\vt}}{\delta_{\vx}} & \norm[\cm]{\nu - \nu_{\Xdes}} \\[2ex]
f(\vx) = \ip[\ch]{K(\cdot,\vx)}{f} & \displaystyle \sup_{f \in \ch : \norm[\ch]{f} \le 1} \abs{\int_{\Omega} f(\vx) \, \varrho(\vx) \, \dif \vx - \frac 1N \sum_{i=1}^N f(\vx_i)}\\[4ex]
K(\vt,\vx) = \cov\bigl(f(\vt), f(\vx) \bigr) & \displaystyle \sqrt{\Ex \abs{\int_{\Omega} f(\vx) \, \varrho(\vx) \, \dif \vx - \frac 1N \sum_{i=1}^N f(\vx_i)}^2}\\\bottomrule
\end{array}
\end{equation*}
\label{tab:ThreeInterpTable}
\end{table}

In the remainder of this section we provide three interpretations of the discrepancy, summarized in Table \ref{tab:ThreeInterpTable}.  These results are presented in various places, including \cites{Hic99a,Hic17a}.  One interpretation of discrepancy is the norm of $\nu - \nu_{\Xdes}$.  The second and third interpretations consider the problem of evaluating the mean of a random variable $Y=f(\vX)$, or equivalently a multidimensional integral
\begin{equation}\label{eq:integration}
\mu = \Ex(Y) = \Ex[f(\vX)] = \int_{\Omega} f(\vx) \, \varrho(\vx) \, \dif \vx,
\end{equation}
where $\vX$ is a random vector with density $\varrho$. The second
interpretation of the discrepancy is worst-case cubature error for integrands, $f$, in the unit ball of a Hilbert space.  The third interpretation is the root mean squared cubature error for integrands, $f$, which are realizations of a stochastic processes.

\subsection{Definition in Terms of a Norm on a Hilbert Space of Measures}
\label{sec:HilbertMeasures}

Let $(\cm, \ip[\cm]{\cdot}{\cdot})$ be a Hilbert space of measures defined on the experimental region, $\Omega$.  Assume that $\cm$ includes all Dirac measures.  Define the kernel function $K:\Omega \times \Omega \to \reals$ in terms of inner products of Dirac measures:
\begin{equation} \label{eq:kerDeltaDef}
    K(\vt,\vx) := \ip[\cm]{\delta_{\vt}}{\delta_{\vx}}, \qquad \forall \vt, \vx \in \Omega.
\end{equation}
The squared distance between two Dirac measures in $\cm$ is then
\begin{equation} \label{eq:distDelta}
   \norm[\cm]{\delta_{\vx} - \delta_{\vt}}^2 = K(\vt,\vt) - 2K(\vt,\vx) + K(\vx,\vx), \qquad \forall \vt, \vx \in \Omega.
\end{equation}

It is straightforward to show that $ K$ is symmetric in its arguments and positive-definite, namely:
\begin{subequations} \label{eq:sympd}
\begin{gather}
K(\vx, \vt) = K(\vt, \vx) \qquad \forall \vt, \vx\in \Omega,\\
\sum\limits_{i, k=1}^N c_i c_k  K(\vx_i,\vx_k) > 0, \qquad \forall N\in\mathbb{N}, \  \vc \in\mathbb{R}^N \setminus \{\vzero\},  \ \Xdes \subset \Omega.
\end{gather}
\end{subequations}
The inner product of arbitrary measures $\lambda, \nu \in \cm$ can be expressed in terms of a double integral of the kernel, $K$:
\begin{equation} \label{eq:ipMdef}
    \ip[\cm]{\lambda}{\nu} = \int_{\Omega \times \Omega} K(\vt,\vx) \, \lambda(\dif \vt) \nu(\dif \vx).
\end{equation}
This can be established directly from \eqref{eq:kerDeltaDef} for $\cm_0$, the vector space spanned by all Dirac measures.  Letting $\cm$ be the closure of the pre-Hilbert space $\cm_0$ then yields \eqref{eq:ipMdef}.

The discrepancy of the design $\Xdes$ with respect to the target probability measure $\nu$ using the kernel $K$ can be defined as the norm of the difference between the target probability measure, $\nu$, and the empirical probability measure for $\Xdes$:
\begin{subequations} \label{eq:discDef}
\begin{align} 
\nonumber
    D^2(\Xdes,\nu,K) & := \norm[\cm]{\nu - \nu_{\Xdes}}^2 \\
    \nonumber
    & = \int_{\Omega \times \Omega} K(\vt,\vx) \, (\nu - \nu_{\Xdes})(\dif \vt) (\nu - \nu_{\Xdes})(\dif \vx) \\
    \nonumber
    & = \int_{\Omega \times \Omega} K(\vt,\vx) \, \nu(\dif \vt) \nu (\dif \vx) - \frac 2N \sum_{i=1}^N \int_{\Omega} K(\vt,\vx_i) \, \nu(\dif \vt)\\
    & \qquad \qquad  + \frac{1}{N^2} \sum_{i,k=1}^N K(\vx_i,\vx_k).
\end{align}
The formula for the discrepancy may be written equivalently in terms of the probability distribution, $F$,  or the probability density, $\varrho$, corresponding to the target probability measure, $\nu$:
\begin{align}
\nonumber
    D^2(\Xdes,F,K)  & = \int_{\Omega \times \Omega} K(\vt,\vx) \, \dif F( \vt) \dif F(\vx) - \frac 2N \sum_{i=1}^N \int_{\Omega} K(\vt,\vx_i)  \, \dif F(\vt) \\ &\qquad \qquad  + \frac{1}{N^2}  \sum_{i,k=1}^N K(\vx_i,\vx_k), \\
\nonumber
   & = \int_{\Omega \times \Omega} K(\vt,\vx) \, \varrho(\vt) \varrho (\vx) \, \dif \vt\dif \vx - \frac 2N \sum_{i=1}^N \int_{\Omega} K(\vt,\vx_i)  \, \varrho(\vt) \, \dif \vt \\
    & \qquad \qquad + \frac{1}{N^2}  \sum_{i,k=1}^N K(\vx_i,\vx_k).
\end{align}
\end{subequations}

Typically the computational cost of evaluating $K(\vt,\vx)$ for any $(\vt,\vx) \in \Omega^2$ is $\Order(d)$, where $\vt$ is a $d$-vector.  Assuming that the integrals above can be evaluated at a cost of $\Order(d)$, the computational cost of evaluating $D(\Xdes,\nu,K)$ is $\Order(dN^2)$.

\begin{figure}
    \centering
    \includegraphics[height = 4cm]{\figdir 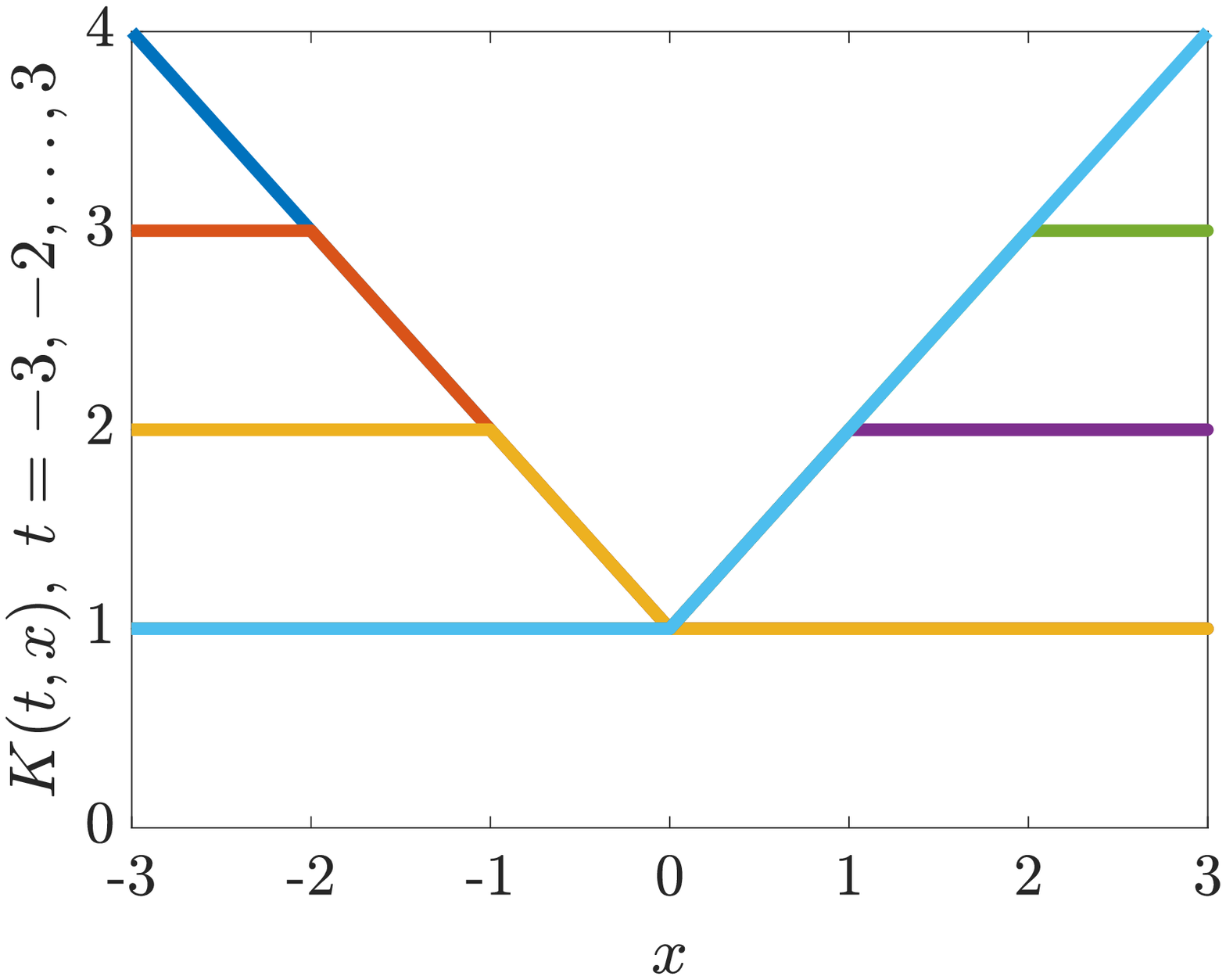} \qquad
    \includegraphics[height = 4cm]{\figdir 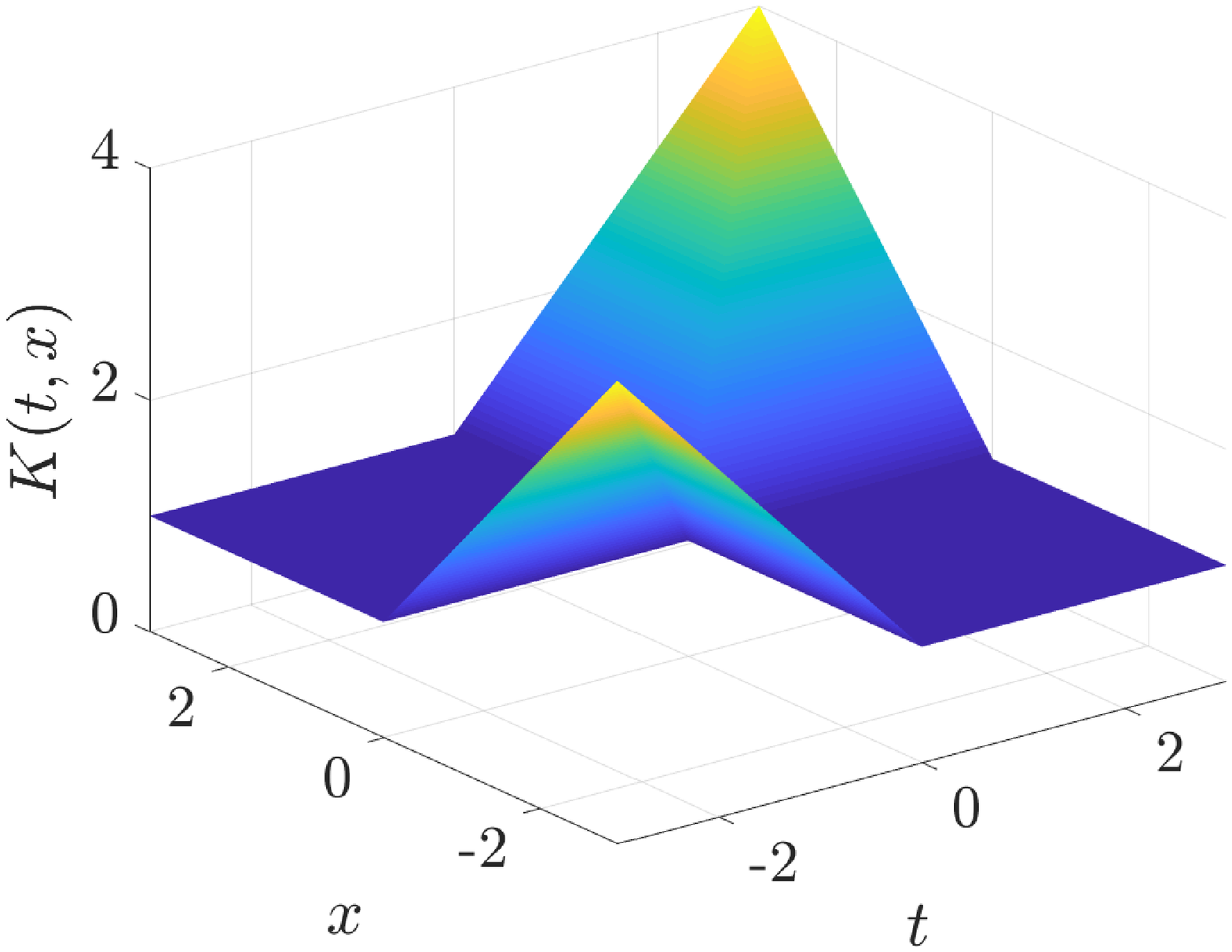}
    \caption{The kernel defined in \eqref{eq:OrigKernel} for $d=1$.}
    \label{fig:kernelpict}
\end{figure}

The formulas for the discrepancy in \eqref{eq:discDef} depend inherently on the choice of the kernel $K$.  That choice is key to answering question \eqref{eq:BigQ}.  An often used kernel is 
\begin{equation} \label{eq:OrigKernel}
     K(\vt,\vx)  = \prod\limits_{j=1}^d\left[1+ \frac 12 \left(|t_j|+ |x_j|- |x_j-t_j| \right)\right].
\end{equation}
This kernel is plotted in Figure \ref{fig:kernelpict} for $d=1$. The distance between two Dirac measures by \eqref{eq:distDelta} for this kernel in one dimension is 
\begin{equation*}
    \norm[\cm]{\delta_{x} - \delta_{t}} = \sqrt{\abs{x-t}}.
\end{equation*}

The discrepancy for the uniform distribution on the unit cube defined in terms of the above kernel is expressed as
\begin{align} 
\nonumber
    D^2(\Udes,F_\unif,K)
    & = \int_{(0,1)^d \times (0,1)^d} K(\vt,\vx) \, \dif \vt \dif \vx - \frac 2N \sum_{i=1}^N \int_{(0,1)^d} K(\vt,\vu_i) \, \dif \vt\\
    \nonumber
    & \qquad \qquad  + \frac{1}{N^2} \sum_{i,k=1}^N K(\vu_i,\vu_k) \\
    \nonumber
    & = \left( \frac 43 \right)^d 
     - \frac 2N \sum_{i=1}^N \prod_{j=1}^d \left[1 + u_{ij} - \frac{u_{ij}^2}{2} \right] \\
     \nonumber
     & \qquad \qquad + \frac 1{N^2} \sum_{i,k=1}^N \prod_{j=1}^d \left[1+ \min(u_{ij},u_{ik})\right].
\end{align}

\subsection{Definition in Terms of a Deterministic Cubature Error Bound}
\label{sec:DetermBound}

Now let $(\ch, \ip[\ch]{\cdot}{\cdot})$ be a reproducing kernel Hilbert space (RKHS) of functions \cite{Aro50}, $f: \Omega \rightarrow \mathbb{R}$, which appear as the integrand in \eqref{eq:integration}. By definition, the reproducing kernel, $K$, is the unique function defined on $\Omega \times \Omega$ with the properties that $K(\cdot, \vx)\in \ch$ for any $\vx \in \Omega$ and $f(\vx)=\ip[\ch]{K(\cdot,\vx)}{f}$.  This second property, implies that $K$ reproduces function values via the inner product. It can be verified that $K$ is symmetric in its arguments and positive definite as in \eqref{eq:sympd}.

The integral $\mu = \int_{\Omega} f(\vx) \, \varrho(\vx) \, \dif \vx$, which was identified as $\Ex[f(\vX)]$ in \eqref{eq:integration}, can be approximated by a sample mean:
\begin{equation}\label{eq:cubature}
\hat{\mu}=\frac{1}{N}\sum_{i=1}^N f(\vx_i).
\end{equation}
The quality of this approximation to the integral, i.e., this cubature, depends in part on how well the empirical distribution of the design, $\Xdes = \{\vx_i\}_{i=1}^N$, matches the target distribution $F$ associated with the density function $\varrho$. 

Define the cubature error as  
\begin{align}
\nonumber 
\err(f,\Xdes) & = \mu - \hat{\mu} = \int_{\Omega} f(\vx)\, \varrho(\vx) \dif \vx -\frac{1}{N}\sum_{i=1}^N f(\vx_i) \\
& =\int_\Omega f(\vx) \, \dif[F(\vx)-F_{\Xdes}(\vx)]. \label{eq:err}
\end{align}
Under modest assumptions on the reproducing kernel,  $\err(\cdot, \Xdes)$ is a bounded, linear functional. 
By the Riesz representation theorem, there exists a unique representer, $\xi \in \ch$, such that 
\[
\err(f, \Xdes)=\ip[\ch]{\xi}{f}, \quad \forall f\in \ch.
\]
The reproducing kernel allows us to write down an explicit formula for that representer, namely, $\xi(\vx)=\ip[\ch]{K(\cdot,\vx)}{\xi}=\ip[\ch]{\xi}{K(\cdot,\vx)}=\err(K(\cdot,\vx),\Xdes)$.
By the Cauchy-Schwarz inequality, there is a tight bound on the squared cubature error, namely
\begin{equation}\label{eq:wcErrA}
\abs{\err(f,\Xdes)}^2=\ip[\ch]{\xi}{f}^2\leq \norm[\ch]{\xi}^2 \norm[\ch]{f}^2 .
\end{equation}
The first term on the right describes the contribution made by the quality of the cubature rule, while the second term describes the contribution to the cubature error made by the nature of the integrand.

The square norm of the representer of the error functional is 
\begin{align*}
\norm[\ch]{\xi}^2& =\ip[\ch]{\xi}{\xi}=\err(\xi,\Xdes) \quad \text{since $\xi$ represents the error functional}\\
& = \err(\err(K(\cdot,\cdot \cdot),\Xdes),\Xdes) \quad \text{since } \xi(\vx)=\err(K(\cdot,\vx),\Xdes)\\
& =\int_{\Omega\times \Omega} K(\vt,\vx)\, \dif[F(\vt)-F_{\Xdes}(\vt)] \dif[F(\vx)-F_{\Xdes}(\vx)] .
\end{align*}
We can equate this formula for $\norm[\ch]{\xi}^2$ with the formula for $D^2(\Xdes,F,K)$ in  \eqref{eq:discDef}.  Thus, the tight, worst-case cubature error  bound in \eqref{eq:wcErrA} can be written in terms of the discrepancy as
\begin{equation*}\label{eq:wcErrB}
\abs{\err(f,\Xdes)} \le \norm[\ch]{f} D(\Xdes,F,K).
\end{equation*}
This implies our second interpretation of the discrepancy in Table \ref{tab:ThreeInterpTable}.

We now identify the RKHS for the kernel $K$ defined in \eqref{eq:OrigKernel}. Let $(\va,\vb)$ be some $d$ dimensional box containing the origin in the interior or on the boundary. For any $\fraku \subseteq \{1, \ldots, d\}$, define $\partial^\fraku f(\vx_\fraku) : = \partial^{|\fraku|}f(\vx_\fraku,\vzero)/\partial \vx_\fraku$, the mixed first-order partial derivative of $f$ with respect to the $x_j$ for $j\in \fraku$, while setting $x_j=0$ for all $j \notin \fraku$. Here, $\vx_{\fraku} = (x_j)_{j \in \fraku}$, and $|\fraku|$ denotes the cardinality of $\fraku$.  By convention, $\partial^{\emptyset}f := f(\vzero)$.  The inner product for the reproducing kernel $K$ defined in  \eqref{eq:OrigKernel} is defined as 
\begin{align}
\label{eq:inner}
\langle f,g \rangle_{\ch} &:= \sum_{\fraku\subseteq \{1,...,d\}}\int_{(\va,\vb)}\partial^{\fraku}f(\vx_\fraku)\partial^{\fraku}g(\vx_\fraku) \, \dif \vx_\fraku \\
\nonumber
& = f(\vzero)g(\vzero)  + \int_{a_1}^{b_1} \partial^{\{1\}}f(x_1)\partial^{\{1\}}g(x_1) \, \dif x_1 \\
\nonumber 
& \qquad + \int_{a_2}^{b_2} \partial^{\{2\}}f(x_2)\partial^{\{2\}}g(x_2) \, \dif x_2 + \cdots \\
\nonumber 
& \qquad + \int_{a_2}^{b_2} \int_{a_1}^{b_1} \partial^{\{1,2\}}f(x_1,x_2)\partial^{\{1,2\}}g(x_1,x_2) \, \dif x_1 \dif x_2+ \cdots \\
\nonumber 
& \qquad + \int_{(\va,\vb)} \partial^{\{1,\ldots, d\}}f(\vx)\partial^{\{1, \ldots, d\}}g(\vx) \, \dif \vx. 
\end{align}

To establish that the inner product defined above corresponds to the reproducing kernel  $K$ defined in \eqref{eq:OrigKernel}, we note that
\begin{align*}
\partial^{\fraku} K((\vx_u,{\bf 0}),\vt) & =\prod_{j \in \fraku}\frac 12 \left[\sign(x_j) - \sign(x_j - t_j) \right] \\
& =
\prod\limits_{j \in \fraku}\sign(t_j) \bbone_{(\min(0,t_j),\max(0,t_j))}(x_j).
\end{align*}
Thus, $K(\cdot,\vt)$ possesses sufficient regularity to have finite $\ch$-norm.  Furthermore, $K$ exhibits the reproducing property for the above inner product because
\begin{align*}
\MoveEqLeft{\langle K(\cdot,\vt),f \rangle}_{\ch} \\
&= \sum_{\fraku\subseteq \{1,...,d\}}\int_{(\va,\vb)} \partial^{\fraku}K((\vx_u,{\bf 0}),\vt)\partial^{\fraku}f(\vx_u,{\bf 0})  \, \dif \vx_u \\
&= \sum_{\fraku\subseteq \{1,...,d\}}\int_{(\va,\vb)}  \prod\limits_{j \in \fraku}\sign(t_j) \bbone_{(\min(0,t_j),\max(0,t_j))}(x_j) \partial^{\fraku}f(\vx_u,{\bf 0})  \, \dif \vx_u \\
& =     \sum_{\fraku\subseteq \{1,...,d\}}\sum_{\frakv\subseteq \fraku}(-1)^{|\fraku|-|\frakv|}f(\vt_\frakv,{\bf 0})= f(\vt).
\end{align*}

\subsection{Definition in Terms of the Root Mean Squared Cubature Error}
\label{sec:RandomBound}

Assume $\Omega$ is a measurable subset in $\mathbb{R}^d$ and $F$ is the target probability distribution defined on $\Omega$ as defined earlier. 
Now, let $f: \Omega \rightarrow \mathbb{R}$ be a stochastic process with a constant pointwise mean, i.e., 
\[
\Ex_{f\in \mathcal{A}}[f(\vx)]=m, \qquad \forall \vx\in \Omega,
\]
where $\mathcal{A}$ is the sample space for this stochastic process. 
Now we interpret $K$ as the \emph{covariance kernel} for the stocastic process:
\[
K(\vt,\vx):=\Ex_{f\in \mathcal{A}}\left( [f(\vt)-m][f(\vx)-m]\right)=\cov(f(\vt),f(\vx)),\qquad \forall \vt, \vx\in \Omega. 
\]
It is straightforward to show that the kernel function is symmetric and positive definite. 

Define the error functional $\err(\cdot,\Xdes)$ in the same way as in \eqref{eq:err}.  Now, the mean squared error is 
\begin{align*}
\Ex_{f\in\mathcal{A}}[(\err(f,\Xdes)]^2 &= \Ex_{f\in\mathcal{A}} \left\{\int_{\Omega}f(\vx) \,\dif F(\vx)-\frac{1}{N}\sum_{i=1}^N f(\vx_i) \right\}^2\\
&=\Ex_{f\in\mathcal{A}}\left\{\int_{\Omega}(f(\vx)-m) \, \dif F(\vx)-\frac{1}{N}\sum_{i=1}^N (f(\vx_i)-m) \right\}^2\\
&=\int_{\Omega^2}\Ex_{f\in \mathcal{A}}[ (f(\vt)-m)(f(\vx)-m)] \, \dif F(\vt)\dif F(\vx)\\
&\qquad  -\frac{2}{N}\sum_{i=1}^N \int_{\Omega} \Ex_{f\in\mathcal{A}}[(f(\vx)-m)(f(\vx_i)-m)]\,\dif F(\vx)\\
&\qquad +\frac{1}{N^2}\sum_{i,k=1}^N\Ex_{f\in\mathcal{A}}[(f(\vx_i)-m)(f(\vx_k)-m)]\\
&=\int_{\Omega^2}K(\vt,\vx) \, \dif F(\vt)\dif F(\vx)-\frac{2}{N}\sum_{i=1}^N \int_{\Omega} K(\vx, \vx_i)\, \dif F(\vx) \\
& \qquad +\frac{1}{N^2} \sum_{i,k=1}^N K(\vx_i,\vx_k). 
\end{align*}
Therefore, we can equate the discrepancy $D(\Xdes, F, K)$ defined in \eqref{eq:discDef} as the root mean squared error:
\[
D(\Xdes, F, K) =\sqrt{\Ex_{f\in\mathcal{A}}[(\err(f,\Xdes)]^2}=\sqrt{\Ex\abs{\int_{\Omega}f(\vx)\varrho(\vx)\dif \vx-\frac{1}{N}\sum_{i=1}^N f(\vx_i)}^2}.
\]

\section{When a Transformed Low Discrepancy Design Also Has Low Discrepancy}
\label{sec:WhenYes}
Having motivated the definition of discrepancy in \eqref{eq:discDef} from three perspectives, we now turn our attention to question \eqref{eq:BigQ}, namely, does a transformation of low discrepancy points with respect to the uniform distribution yield low discrepancy points with respect to the new target distribution. In this section, we show a positive result, yet recognize some qualifications.

Consider some symmetric, positive definite kernel, $K_\unif : (0,1)^d \times (0,1)^d \to \reals$, some uniform design $\Udes$, some other domain, $\Omega$, some other target distribution, $F$, and some transformation $\vPsi:(0,1)^d \to \Omega$ as defined in \eqref{eq:transPts}. Then the squared discrepancy of the uniform design can be expressed according to \eqref{eq:discDef} as follows:
\begin{align} 
\nonumber
    \MoveEqLeft{D^2(\Udes,F_{\unif},K_\unif)} \\
    \nonumber
    & = \int_{(0,1)^d \times (0,1)^d} K_\unif(\vu,\vv) \,  \dif \vu \dif \vv - \frac 2N \sum_{i=1}^N \int_{\Omega} K_\unif(\vu,\vu_i) \, \dif \vu\\
    \nonumber
    & \qquad \qquad  + \frac{1}{N^2} \sum_{i,k=1}^N K_\unif(\vu_i,\vu_k) \\
    \nonumber
    & = \int_{\Omega \times \Omega} K_\unif(\vPsi^{-1}(\vt),\vPsi^{-1}(\vx)) \, \abs{\frac{\partial \vPsi^{-1}(\vt)}{\partial \vt}} \abs{\frac{\partial \vPsi^{-1}(\vx)}{\partial \vx}} \,\dif \vt \dif \vx \\
    \nonumber 
    & \qquad \qquad  - \frac 2N \sum_{i=1}^N \int_{\Omega} K_\unif(\vPsi^{-1}(\vt),\vPsi^{-1}(\vx_i)) \, \abs{\frac{\partial \vPsi^{-1}(\vt)}{\partial \vt}} \, \dif \vt\\
    \nonumber
    & \qquad \qquad  + \frac{1}{N^2} \sum_{i,k=1}^N K_\unif(\vPsi^{-1}(\vx_i),\vPsi^{-1}(\vx_k)) \\
    \nonumber
    & = D^2(\Xdes,F,K)
\end{align}
where the kernel $K$ is defined as
\begin{subequations} \label{eq:BigQYes}
\begin{equation} \label{eq:BigQYesa}
    K(\vt,\vx) = K_{\unif}(\vPsi^{-1}(\vt),\vPsi^{-1}(\vx)),
\end{equation}
\emph{and provided that} the density, $\varrho$, corresponding to the target distribution, $F$, satisfies
\begin{equation} \label{eq:BigQYesb}
    \varrho(\vx) = \abs{\frac{\partial \vPsi^{-1}(\vx)}{\partial \vx}}.
\end{equation}
\end{subequations}
The above argument is summarized in the following theorem.

\begin{theorem}
Suppose that the design $\Xdes$ is constructed by transforming the design $\Udes$ according to the transformation \eqref{eq:transPts}.  Also suppose that conditions \eqref{eq:BigQYes} are satisfied.  Then $\Xdes$ has the same discrepancy with respect to the target distribution, $F$, defined by the kernel $K$ as does the original design $\Udes$ with respect to the uniform distribution and defined by the kernel $K_\unif$.  That is,
\begin{equation*}
     D(\Xdes,F,K) = D(\Udes,F_{\unif},K_\unif).
\end{equation*}
As a consequence, under conditions \eqref{eq:BigQYes}, question \eqref{eq:BigQ} has a positive answer.
\end{theorem}

Condition \eqref{eq:BigQYesb} may be easily satisfied.  For example, it is automatically satisfied by the inverse cumulative distribution transform \eqref{eq:inverse}.  Condition \eqref{eq:BigQYesa} is simply a matter of definition of the kernel, $K$, but this definition has consequences.  From the perspective of Section \ref{sec:HilbertMeasures}, changing the kernel from $K_\unif$ to $K$ means changing the definition of the distance between two Dirac measures.  From the perspective of Section \ref{sec:DetermBound}, changing the kernel from $K_\unif$ to $K$ means changing the definition of the Hilbert space of integrands, $f$, in \eqref{eq:integration}.   From the perspective of Section \ref{sec:RandomBound}, changing the kernel from $K_\unif$ to $K$ means changing the definition of the covariance kernel for the integrands, $f$, in \eqref{eq:integration}.

To illustrate this point, consider a cousin of the kernel in \eqref{eq:OrigKernel}, which places the reference point at $\boldsymbol{0.5} = (0.5, \ldots, 0.5)$, the center of the unit cube $(0,1)^d$:
\begin{align}
\label{eq:L2kernel}
    K_{\unif}(\vu,\vv) &= \prod_{j=1}^d\left[1+\frac{1}{2}\left(\left|u_j-1/2\right|+ \left|v_j- 1/2 \right|-\left |u_j-v_j \right| \right) \right] \\
    \nonumber
    & = K(\vu - \boldsymbol{0.5}, \vv - \boldsymbol{0.5}) \qquad \text{for $K$ defined in \eqref{eq:OrigKernel}}. 
\end{align}
This kernel defines the centered $L^2$-discrepancy \cite{Hic97a}.
Consider the standard multivariate normal distribution, $F_\normal$, and choose the inverse normal distribution,
\begin{equation} \label{eq:invNormTrans}
    \vPsi(\vu) = (\Phi^{-1}(u_1), \ldots, \Phi^{-1}(u_d)),
\end{equation}
where $\Phi$ denotes the standard normal distribution function.  Then condition \eqref{eq:BigQYesb} is automatically satisfied, and condition \eqref{eq:BigQYesa} is  satisfied by defining
\begin{align}
\nonumber
     K(\vt,\vx) &= K_{\unif}(\vPsi^{-1}(\vt),\vPsi^{-1}(\vx)) \\
     \nonumber
      &= \prod_{j=1}^d\left[1+\frac{1}{2}\left(\left|\Phi(t_j)-1/2\right|+ \left|\Phi(x_j)- 1/2 \right| \right . \right . \\
      \nonumber
      & \qquad \qquad \left.  \left . -\left|\Phi(t_j)-\Phi(x_j)\right|\right)\right].
\end{align}
In one dimension, the distance between two Dirac measures defined using the kernel $K_\unif$ above is $\norm{\delta_{x} - \delta_{t}} = \sqrt{|x-t|}$, whereas the distance defined using the kernel $K$ above is $\norm{\delta_{x} - \delta_{t}} = \sqrt{|\Phi(x)-\Phi(t)|}$. Under kernel $K$, the distance between two Dirac measures is bounded, even though the domain of the distribution is unbounded.  Such an assumption may be unpalatable.

\section{Do Transformed Low Discrepancy Points Have Low Discrepancy More Generally} \label{sec:DoTransformedPts}

The discussion above indicates that condition \eqref{eq:BigQYesa} can be too restrictive.  We would like to compare the discrepancies of designs under kernels that do not satisfy that restriction.  In particular, we consider the centered $L^2$-discrepancy for uniform designs on $(0,1)^d$ defined by the kernel in \eqref{eq:L2kernel}:
\begin{align*}
\MoveEqLeft{D^2(\Udes, F_\unif, K_\unif)} \\
& = \left(\frac{13}{12}\right)^d - \frac{2}{N}\sum_{i=1}^N \prod_{j=1}^d \left[1+\frac{1}{2}\left(|u_{ij}-1/2|-|u_{ij}-1/2|^2\right) \right]\\
& \qquad + \frac{1}{N^2}\sum_{i,k=1}^N\prod_{j=1}^d\left[1+\frac{1}{2}\left(|u_{ij}-1/2|+|u_{kj}-1/2|-|u_{ij}-u_{kj}| \right) \right],  
\end{align*}
where again, $F_\unif$ denotes the uniform distribution on $(0,1)^d$, and $\Udes$ denotes a design on $(0,1)^d$

Changing perspectives slightly, if $F_\unif'$ denotes the uniform distribution on the cube of volume one centered at the origin, $(-0.5,0.5)^d$, and the design $\Udes'$ is constructed by  subtracting $\boldsymbol{0.5}$ from each point in the design $\Udes$:
\begin{equation} \label{eq:Updef}
    \Udes' = \{\vu -\boldsymbol{0.5} : \vu \in \Udes \},
\end{equation}
then 
\begin{equation*} \label{eq:desEquiv}
     D(\Udes',F'_\unif,K) = D(\Udes, F_\unif, K_\unif),
\end{equation*}
where $K$ is the kernel defined in \eqref{eq:OrigKernel}.  

Recall that the origin is a special point in the definition of the inner product for the Hilbert space with $K$ as its reproducing kernel in \eqref{eq:inner}.  Therefore, this $K$ from \eqref{eq:OrigKernel} is appropriate for defining the discrepancy for target distributions centered at the origin, such as the standard normal distribution, $F_\normal$.  Such a discrepancy is 
\begin{align}
\nonumber
\MoveEqLeft{D^2(\Xdes, F_\normal, K) = \left(1+\sqrt{\frac{2}{\pi}}\right)^d}\\
\nonumber
  & - \frac{2}{N}\sum_{i=1}^N \prod\limits_{j=1}^d\left[ 1+\frac{1}{\sqrt{2\pi}}+\frac{1}{2}|x_{ij}|-x_{ij}\left(\Phi(x_{ij})-\frac 12 \right)-\phi(x_{ij})\right]\\
  &+\frac{1}{N^2}\sum_{i,k=1}^N \prod_{j=1}^d \left[1+\frac{1}{2}\left(|x_{ij}|+|x_{kj}|-|x_{ij}-x_{kj}|\right)\right]. 
  \label{eq:DisNormald>1}
\end{align}
Here, $\phi$ is the standard normal probability density function. 
The derivation of \eqref{eq:DisNormald>1} is given in the Appendix. 

We numerically compare the discrepancy of a uniform design, $\Udes'$ given by \eqref{eq:Updef} and the discrepancy of a design constructed by the inverse normal transformation, i.e.,  $\Xdes = \vPsi(\Udes)$ for $\vPsi$ in \eqref{eq:invNormTrans}, where the $\Udes$ leading to both $\Udes'$ and $\Xdes$ is identical.  We do not expect the magnitudes of the discrepancies to be the same, but we ask
\begin{multline} \tag{Q'} \label{eq:BigQPrime}
    \text{Does } D(\Udes'_1,F'_\unif, K)\leq D(\Udes_2',F_\unif', K) \\
    \text{imply }
     D(\vPsi(\Udes_1),F_\normal, K)\leq D(\vPsi(\Udes_2),F_\normal, K)?
\end{multline}
Again, $K$ is given by \eqref{eq:OrigKernel}.  So we are actually comparing discrepancies defined by the same kernels, but \emph{not kernels that satisfy \eqref{eq:BigQYesa}}.

\begin{figure}[ht]
\begin{center}
\includegraphics[width=8cm]{\figdir 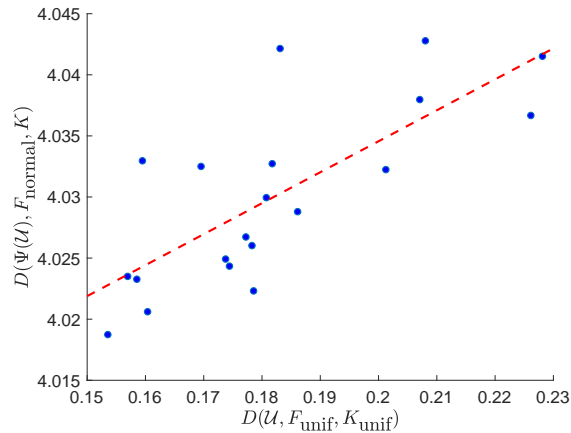}
\caption{Normal discrepancy versus uniform discrepancy for transformed designs. \label{fig:UniVsNormDisc}}
\end{center}
\end{figure}

Let $d=5$ and $N=50$. 
We generate $B=20$ independent and identically distributed (IID) uniform designs, $\Udes$ with $N=50$ points on $(0,1)^5$ and then use the inverse distribution transformation to obtain IID random $N({\bf 0}, {\mathsf I}_5)$ designs, $\Xdes = \vPsi(\Udes)$. 
Figure \ref{fig:UniVsNormDisc} plots the discrepancies for normal designs, $D(\vPsi(\Udes),F_\normal, K)$,  against the discrepancies for the uniform designs, $D(\Udes,F_\unif,K_\unif) = D(\Udes',F_\unif',K)$ for each of the $B=20$ designs. 
Question \eqref{eq:BigQPrime} has a positive answer if and only if the lines passing through any two points on this plot all have non-negative slopes.  However, that is not the case.  Thus \eqref{eq:BigQPrime} has a negative answer.

\begin{figure}[ht]
\begin{center}
\includegraphics[width=8cm]{\figdir 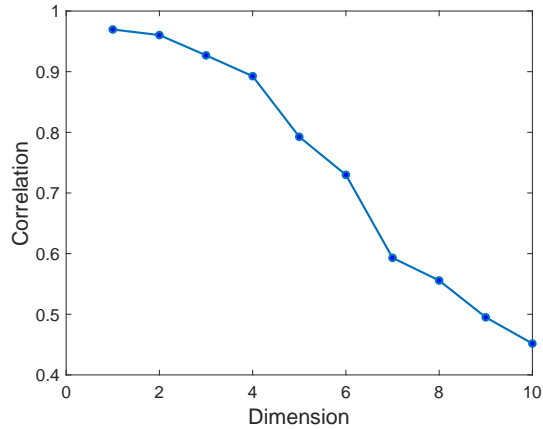}
\caption{Correlation between the uniform and normal discrepancies for different dimensions. \label{fig:DiscVsd}}
\end{center}
\end{figure}

We further investigate the relationship between the discrepancy of a uniform design and the discrepancy of the same design after inverse normal transformation.
Varying the dimension $d$ from $1$ to $10$, we calculate the sample correlation between $D(\vPsi(\Udes),F_\normal, K)$ and $D(\Udes,F_\unif,K_\unif) = D(\Udes',F_\unif',K)$ for $B=500$ IID designs of size $N=50$.  Figure \ref{fig:DiscVsd} displays the correlation as a function of $d$. Although the correlation is positive, it degrades with increasing $d$.

\begin{example} A simple cubature example illustrates that an inverse transformed low discrepancy design, $\Udes$, may yield a large $D(\vPsi(\Udes),F_\normal, K)$ and also a large cubature error. Consider the integration problem in \eqref{eq:integration} with
\begin{subequations} \label{eq:badexample}
\begin{gather}
\vX \sim  N(\vzero, {\mathsf I}_d), \qquad f(\vx) = \frac{x_1^2+\cdots+x^2_d}{1+10^{-8}(x_1^2+\cdots+x_d^2)}, \qquad Y = f(\vX),\\
\label{eq:badexampleb}
\mu = \Ex(Y) = \int_{\reals^d} \frac{x_1^2+\cdots+x^2_d}{1+10^{-8}(x_1^2+\cdots+x_d^2)} \phi(\vx) \, \dif \vx,
\end{gather}
\end{subequations}
where $\phi$ is the probability density function for the standard multivariate normal distribution.  The function $f:\reals^d \to \reals$ is constructed to asymptote to a constant as $\norm[2]{\vx}$ tends to infinity to ensure that $f$ lies inside the Hilbert space corresponding to the kernel $K$ defined in \eqref{eq:OrigKernel}.
Since the integrand in \eqref{eq:badexample} is a function of $\norm[2]{\vx}$, $\mu$ can be written as a one dimensional integral.  For $d=10$, $\mu = 10$ to at least $15$ significant digits using quadrature.

We can also approximate the integral in \eqref{eq:badexample} using a $d=10$, $N=512$ cubature \eqref{eq:cubature}.  We compare cubatures using two designs.  The design $\Xdes_1$ is the inverse normal transformation of a scrambled Sobol' sequence, $\Udes_1$, which has a low discrepancy with respect to the uniform distribution on the $d$-dimensional unit cube.  The design $\Udes_2$ takes the point in $\Udes_1$ that is closet to $\vzero$ and moves it to $\left(10^{-15}, \ldots, 10^{-15}\right)$, which is very close to $\vzero$.  As seen in Table \ref{tab:badexample}, the two uniform designs have quite similar, small discrepancies.  However, the transformed designs, $\Xdes_j = \vPsi(\Udes_j)$ for $j=1,2$, have much different discrepancies with respect to the normal distribution.  This is due to the point in $\Xdes_2$ that has large negative coordinates.  Furthermore, the cubatures, $\hat{\mu}$, based on these two designs have significantly different errors.  The first design has both a smaller discrepancy and a smaller cubature error than the second.  This could not have been inferred by looking at the discrepancies of the original uniform designs.

\begin{table}[htbp]
  \centering
  \caption{Comparison of Integral Estimate}
    \begin{tabular}{ccccc}
   $\Udes$       & $D(\Udes,F_\unif, K)$     & $D(\vPsi(\Udes),F_\normal, K)$     & $\hat{\mu}$  & Relative Error\\
    \toprule
    $\Udes_1$     &   0.0285    &  18.57     & 10.0182 & 0.0018\\
    $\Udes_2$     &  0.0292      &  58.82     & 11.2238 & 0.1224
    \end{tabular}%
  \label{tab:badexample}%
\end{table}%


\end{example}

\section{Improvement by the Coordinate-Exchange Method}\label{sec:CoordEx}

In this section, we propose an efficient algorithm that improves a design's quality in terms of the discrepancy for the target distribution. 
We start with a low discrepancy uniform design, such as a Sobol' sequence, and transform it into a design that approximates the target distribution. 
Following the optimal design approach, we then apply a coordinate-exchange algorithm to further improve the discrepancy of the design.

The coordinate-exchange algorithm was introduced in \cite{meyer1995coordinate}, and then applied widely to construct various kinds of optimal designs \cites{sambo2014coordinate,overstall2017bayesian,kang2018stochastic}. 
The coordinate-exchange algorithm is an iterative method.
It finds the ``worst" coordinate $x_{ij}$ of the current design and replaces it to decrease loss function, in this case, the discrepancy. 
The most appealing advantage of the coordinate-exchange algorithm is that at each step one need only solve a univariate optimization problem.

First, we define the point deletion function, $\frakd_p$, as the change in square discrepancy resulting from removing the a point from the design:
\begin{equation}\label{eq:rowdeletion}
\frakd_p(i) = D^2(\Xdes)-\left(\frac{N-1}{N}\right)^2 D^2(\Xdes\backslash \{\vx_i\}).
\end{equation}
Here, the design $\Xdes\backslash \{\vx_i\}$ is the $N-1$ point design with the point $\{\vx_i\}$ removed. We suppress the choice of target distribution and kernel in the above discrepancy notation for simplicity. We then choose 
\begin{equation*} \label{eq:rowchoice}
    i^*=\argmax_{i=1,\ldots,N} \frakd_p(i).
\end{equation*}
The definition of $i^*$ means that removing $\vx_{i^*}$ from the design $\Xdes$ results in the smallest discrepancy among all possible deletions.  Thus, $\vx_{i^*}$ is helping the least, which makes it a prime candidate for modification.

Next, we define a coordinate deletion function, $\frakd_{c}$, as the change in the square discrepancy resulting from removing a coordinate in our calculation of the discrepancy:
\begin{equation}\label{eq:coorddeletion}
\frakd_c(j) = D^2(\Xdes)-D^2(\Xdes_{-j}).
\end{equation}
Here, the design $\Xdes_{-j}$ still has $N$ points but now only $d$ dimensions, the $j^{\text{th}}$ coordinate having been removed.  For this calculation to be feasible, the target distribution must have independent marginals.  Also, the kernel must be of product form.  To simplify the derivation, we assume a somewhat stronger condition, namely that the marginals are identical and that each term in the product defining the kernel is the same for every coordinate:
\begin{equation}
    \label{eq:prodkernel}
    \Omega = \tOmega \times \cdots \times \tOmega, \qquad K(\vt,\vx) = \prod_{j=1}^d [1+ \tK(t_j,x_j)], \qquad \tK:\tOmega \times \tOmega \to \reals.
\end{equation}
We then choose 
\begin{equation*} \label{eq:columnchoice}
    j^*=\argmax_{j=1, \ldots, d} \frakd_c(j).
\end{equation*}
For reasons analogous to those given above, the $j^{*\text{th}}$ coordinate seems to be the best candidate for change.

Let $\Xdes^*(x)$ denote the design that results from replacing $x_{i^*j^*}$ by $x$. 
We now define $\Delta(x)$ as improvement in the squared discrepancy resulting from replacing $\Xdes$ by $\Xdes^*(x)$:
\begin{equation}\label{eq:deltafun}
\Delta(x) = D^2(\Xdes)-D^2(\Xdes^*(x)).
\end{equation}
We can reduce the discrepancy by find an $x$ such that $\Delta(x)$ is positive. 
The coordinate-exchange algorithm outlined in Algorithm \ref{alg:coorex} improves the design by maximizing $\Delta(x)$ for one chosen coordinate in one iteration. 
The algorithm terminates when it exhausts the maximum allowed number of iterations or the optimal improvement $\Delta(x^*)$ is so small that it becomes negligible ($\Delta(x^*)\leq \textrm{TOL}$). 
Algorithm \ref{alg:coorex} is a greedy algorithm, and thus it can stop at a local optimal design. 
We recommend multiple runs of the algorithm with different initial designs to obtain a design with the lowest discrepancy possible. 
Alternatively, users can include stochasticity in the choice of the coordinate that is to be exchanged, similarly to \cite{kang2018stochastic}. 

\begin{algorithm}[ht]
\caption{Coordinate Exchange Algorithm.\label{alg:coorex}}
\begin{algorithmic}[1]
\INPUT An initial design $\Xdes$ on the domain $\Omega$, a target distribution, $F$, a kernel, $K$ of the form \eqref{eq:prodkernel}, a small value $\TOL$ to determine the convergence of the algorithm, and the maximum allowed number of iterations, $M_{\max}$.
\OUTPUT Low discrepancy design $\Xdes$. 
\For{$m=1,2,\ldots, M_{\max}$}
\State Compute the point deletion function $\frakd_p(1), \ldots, \frakd_p(N)$. Choose the $i^*$-th point which has the largest point deletion value, i.e. $i^* = \argmax_{i} \frakd_p(i)$. 
\State Compute the coordinate deletion function $\frakd_c(1), \ldots, \frakd_c(d)$ and choose the $j^*$-th coordinate which has the largest coordinate deletion value, i.e., $j^* = \argmax_{j} \frakd_c(j)$. 
\State Replace the coordinate $x_{i^*j^*}$ by $x^*$ which is defined by the univariate optimization problem
\[
x^* = \argmax_{x \in \tOmega} \Delta(x).
\]
\If{$\Delta(x^*)>\TOL$} 
\State Replace $x_{i^*j^*}$ with $x^*$ in the design $\Xdes$, i.e., let  $\Xdes(x^*)$ replace the old $\Xdes$.
\Else 
\State Terminate the loop.

\EndIf
\EndFor
\State Return the design, $\Xdes$, and the discrepancy, $D(\Xdes,F,K)$.
\end{algorithmic}
\end{algorithm}

For kernels of product form, \eqref{eq:prodkernel}, and target distributions with independent and identical marginals, the formula for the squared discrepancy in \eqref{eq:discDef} becomes
\begin{subequations}
\label{eq:chHdef}
\begin{align}
\nonumber
    D^2(\Xdes,\rho,K)  & = (1+c)^d  - \frac 2N \sum_{i=1}^N H(\vx_{i}) + \frac{1}{N^2}  \sum_{i,k=1}^N K(\vx_{i},\vx_{k}),
    \intertext{where}
    h(x) & = \int_{\tOmega} \tK(t,x)  \, \tvarrho(t) \, \dif t, \\
    c & = \int_{\tOmega \times \tOmega} \tK(t_k,x_k) \, \tvarrho(t) \tvarrho (x) \, \dif t\dif x = \int_{\tOmega} h(x) \, \tvarrho(x) \, \dif x, \\
    H(\vx) &= \prod_{j=1}^d [1+ h(x_{j})].
\end{align}
\end{subequations}
An evaluation of $h(x)$ and $\tK(t,x)$ each require $\Order(1)$ operations, while an evaluation of $H(\vx)$ and $K(\vt,\vx)$ each require $\Order(d)$ operations.  The computation of $D(\Xdes,\rho,K)$ requires $\Order(dN^2)$ operations because of the double sum.
For a standard multivariate normal target distribution and the kernel defined in \eqref{eq:OrigKernel}, we have
\begin{align*}
c &= \sqrt{\frac{2}{\pi}}, \\
h(x) &= \frac{1}{\sqrt{2\pi}}+\frac{1}{2}|x|-x[\Phi(x)-1/2]-\phi(x),\\
\tK(t,x) &= \frac{1}{2} (|t|+ |x|- |x-t|).
\end{align*}

The point deletion function defined in \eqref{eq:rowdeletion} then can be expressed as
\begin{align}
\nonumber
\frakd_p(i) 
&= \frac{(2N-1)(1+c)^d}{N^2} - \frac{2}{N}\biggl[ \frac{1}{N} \sum_{k=1}^N  H(\vx_{k}) + \left(1 - \frac 1N \right) H(\vx_{i})  \biggr] \\
\nonumber
& \qquad \qquad + \frac{1}{N^2}\biggl[2 \sum_{k=1}^N K(\vx_{i},\vx_{j})- K(\vx_{i},\vx_{i})\biggr]. 
\end{align}
The computational cost for $\frakd_p(1), \ldots, \frakd_p(N)$ is then $\Order(dN^2)$, which is the same order as the cost of the discrepancy of a single design.

The coordinate deletion function defined in \eqref{eq:coorddeletion} can be be expressed as
\begin{multline*}
\frakd_c(j) = (c-1)c^{d-1} -\frac{2}{N}\sum_{i=1}^N \frac{h(x_{ij})H(\vx_i)}{1+h(x_{ij})}  \\
+\frac{1}{N^2}\sum_{i,k=1}^N \frac{\tK(x_{ij},x_{kj}) K(\vx_i,\vx_j)}{1+\tK(x_{ij},x_{kj})}  .
\end{multline*}
The computational cost for $\frakd_c(1), \ldots, \frakd_p(d)$ is also $\Order(dN^2)$, which is the same order as the cost of the discrepancy of a single design.

Finally, the  function $\Delta$ defined in \eqref{eq:deltafun} is given by
\begin{align}\label{eq:deltafunction}
\nonumber
\Delta(x)
&= -\frac{2\left[  h(x_{i^*j^*})-h(x)    \right]H(\vx_{i^*})}{N[1 + h(x_{i^*j^*})]}   \\
\nonumber
& \qquad \qquad +\frac{1}{N^2}\left(2\sum_{\substack{i=1 \\ i \ne i^*}}^N  \frac{[\tK(x_{i^*j^*},x_{ij^*})-\tK(x,x_{ij^*})] K(\vx_{i^*},\vx_i)} {1 + \tK(x_{i^*j^*},x_{ij^*})} \right . \\
\nonumber
& \qquad \qquad \left . + \frac{[\tK(x_{i^*j^*},x_{i^*j^*})-\tK(x,x)] K(\vx_{i^*},\vx_{i^*})} {1 + \tK(x_{i^*j^*},x_{i^*j^*})} \right )
\end{align} 
If we drop the terms that are independent of $x$, then we can maximize the function
\begin{equation*} 
\Delta'(x) = Ah(x)  - \frac{1}{N}\sum_{\substack{i=1 \\ i \ne i^*}}^N B_i \tK(x,x_{ij^*}) - C \tK(x,x)
\end{equation*}
where
\begin{equation*}
A = \frac{2H(\vx_{i^*})}{1 + h(x_{i^*j^*})}, \quad
B_i  = \frac{2K(\vx_{i^*},\vx_i)} {1 + \tK(x_{i^*j^*},x_{ij^*})}, \quad
C  = \frac{K(\vx_{i^*},\vx_{i^*})} {N[1 + \tK(x_{i^*j^*},x_{i^*j^*})]}.
\end{equation*}
Note that $A, B_1, \ldots, B_N, C$ only need to be computed once for each iteration of the coordinate exchange algorithm.

Note that the coordinate-exchange algorithm we have developed is a greedy and deterministic algorithm. 
The coordinate that we choose to make exchange is the one has the largest point and coordinate deletion function values, and we always make the exchange for new coordinate as long as the new optimal coordinate improves the objective function. 
It is true that such deterministic and greedy algorithm is likely to return a design of whose discrepancy attains a \emph{local} minimum. 
To overcome this, we can either run the algorithm with multiple random initial designs, or we can combine the coordinate-exchange with stochastic optimization algorithms, such as simulated annealing (SA) \cite{kirkpatrick1983optimization} or threshold accepting (TA) \cite{fang2003lower}. 
For example, we can add a random selection scheme when choosing a coordinate to exchange, and when making the exchange of the coordinates, we can incorporate a random decision to accept the exchange or not. 
The random decision can follow the SA or TA method. 
Tuning parameters must be carefully chosen to make the SA or TA method effective. 
Interested readers can refer to \cite{winker1997application} to see how TA can be applied to the minimization of discrepancy. 

\section{Simulation} \label{sec:Simulation}

To demonstrate the performance of the $d$-dimensional standard normal design proposed in Section \ref{sec:CoordEx}, we compare three families of designs: (1) RAND: inverse transformed IID uniform random numbers;  (2) SOBOL: inverse transformed Sobol' set; (3) E-SOBOL: inverse transformed scrambled Sobol' set where the one dimensional projections of the Sobol' set have been adjusted to be  $\left\{1/(2N), 3/(2N), \ldots, (2N-1)/(2N) \right\}$; and (4) CE: improved E-SOBOL via Algorithm \ref{alg:coorex}. 
We have tried different combinations of dimension, $d$, and sample size, $N$. 
For each $(d,N)$ and each algorithm we generate $500$ designs and compute their discrepancies \eqref{eq:DisNormald>1}.  

\begin{figure}[ht]
\begin{center}
\includegraphics[width=8cm]{\figdir 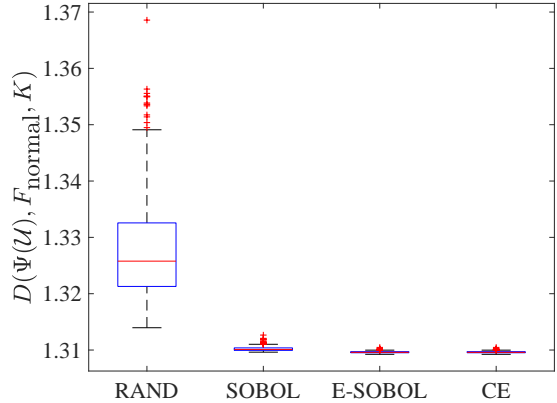}
\caption{Performance comparison of designs. \label{fig:SOBOLVsRAND}}
\end{center}
\end{figure}

\begin{figure}[ht]
\centering
{\subfigure[$d=2,N=32$] {\includegraphics[width=5.5cm]{\figdir 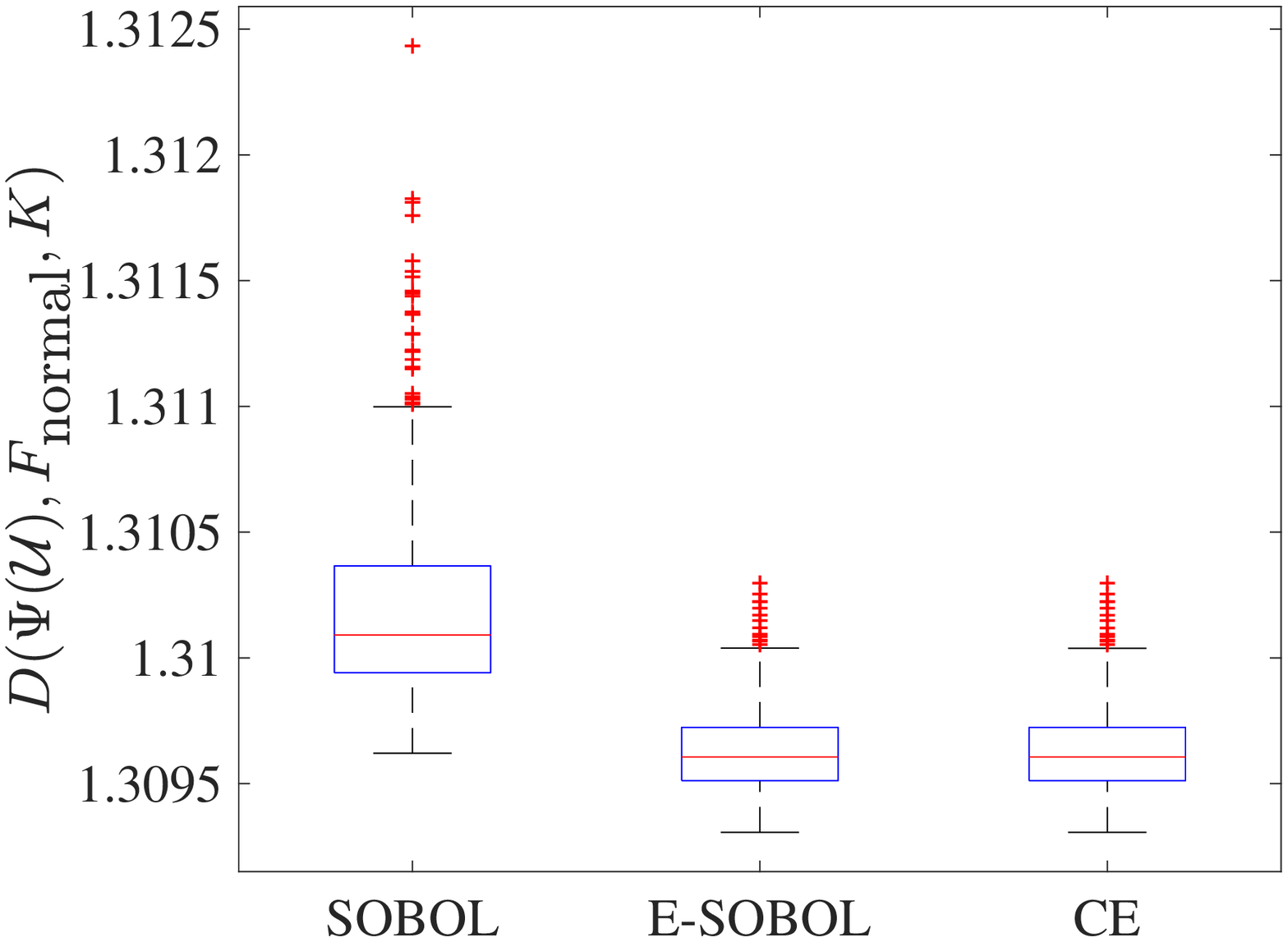}}}
\quad
{\subfigure [$d=3, N=64$]
{\includegraphics[width=5.5cm]{\figdir 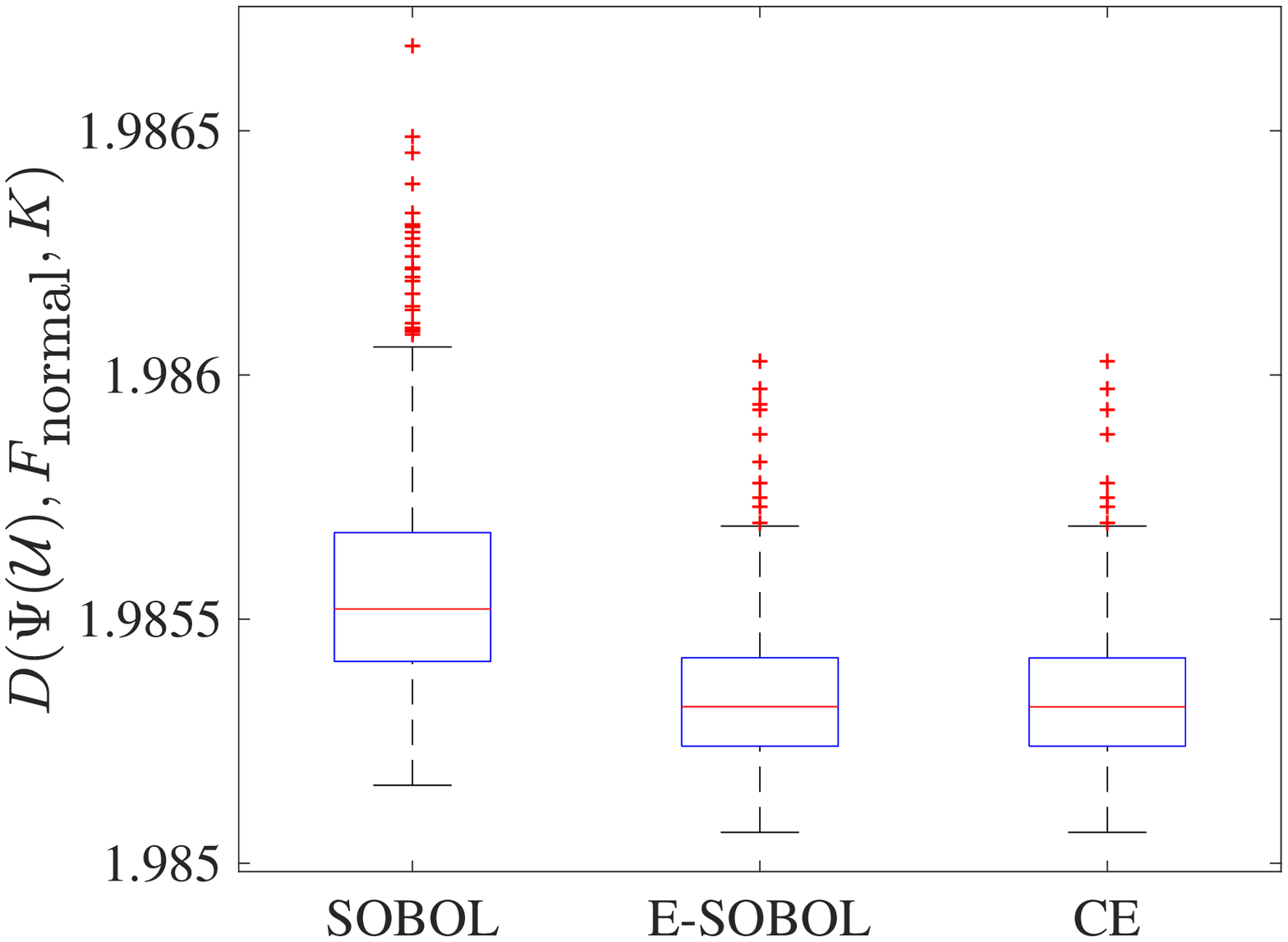}}}
\quad
{\subfigure[$d=4, N=64$] {\includegraphics[width=5.5cm]{\figdir 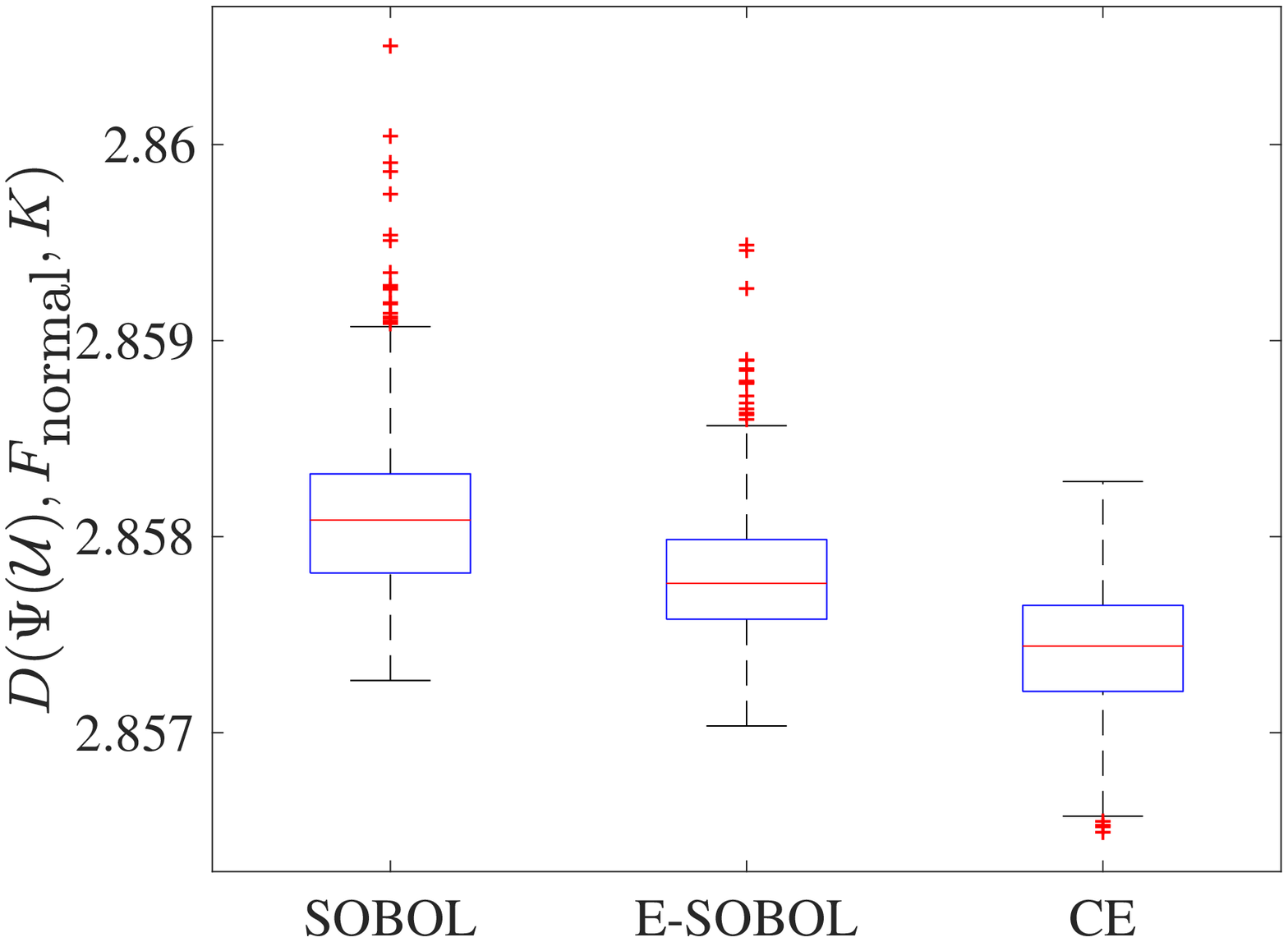}}}
\quad
{\subfigure[$d=6, N=128$] {\includegraphics[width=5.5cm]{\figdir 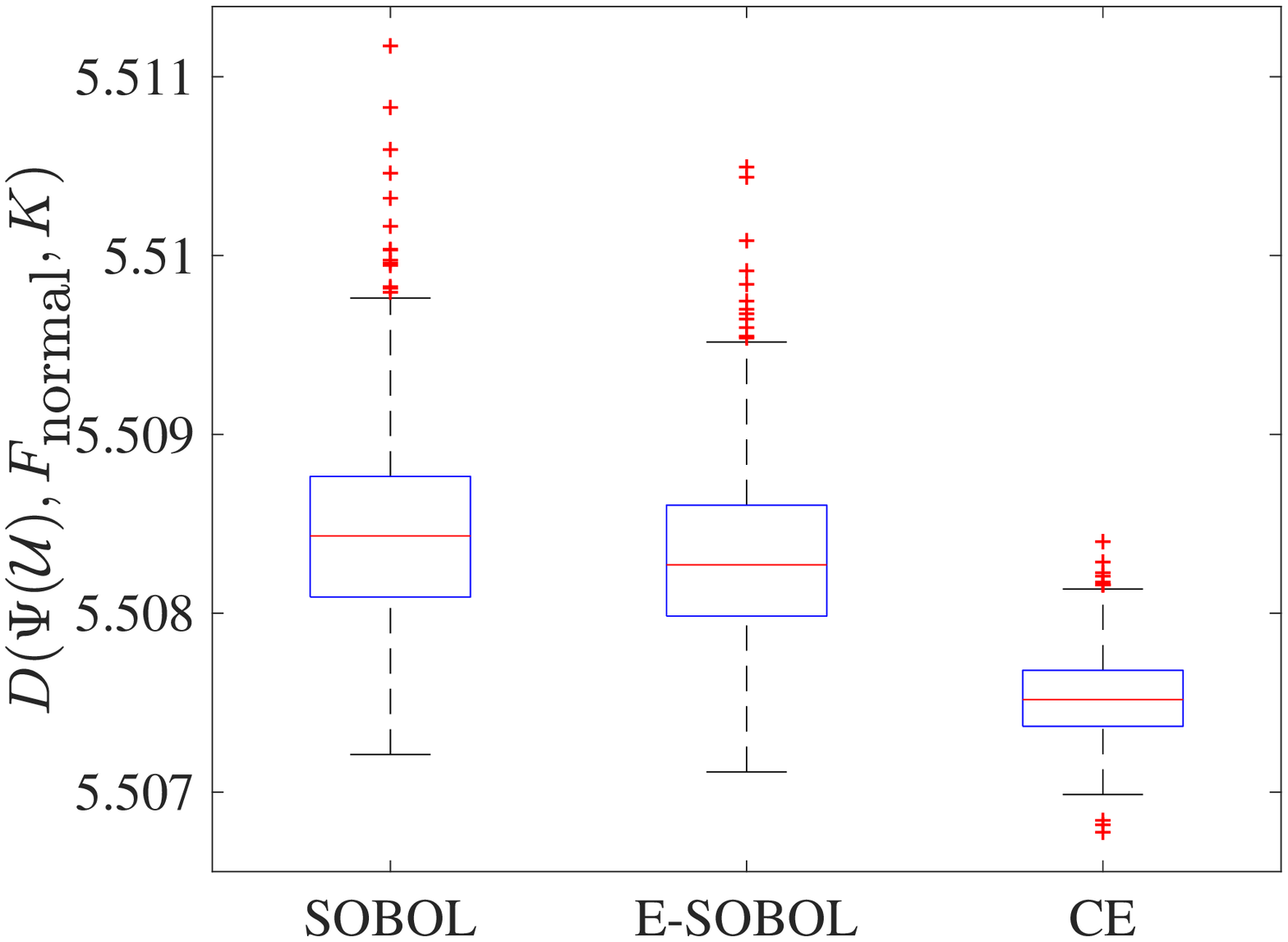}}}
\quad
{\subfigure[$d=8, N=256$] {\includegraphics[width=5.5cm]{\figdir 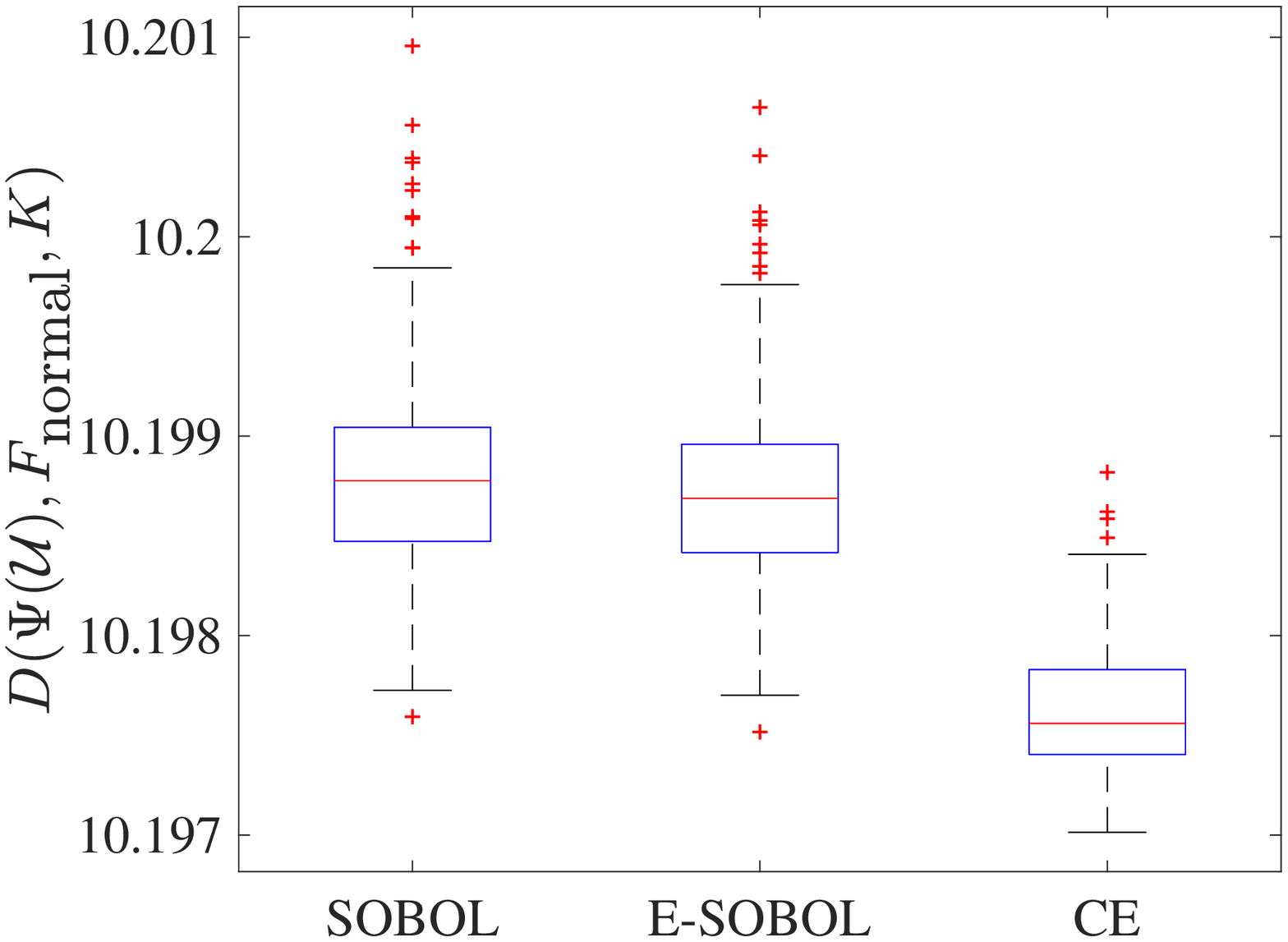}}}
\quad
{\subfigure[$d=10, N=512$] {\includegraphics[width=5.5cm]{\figdir 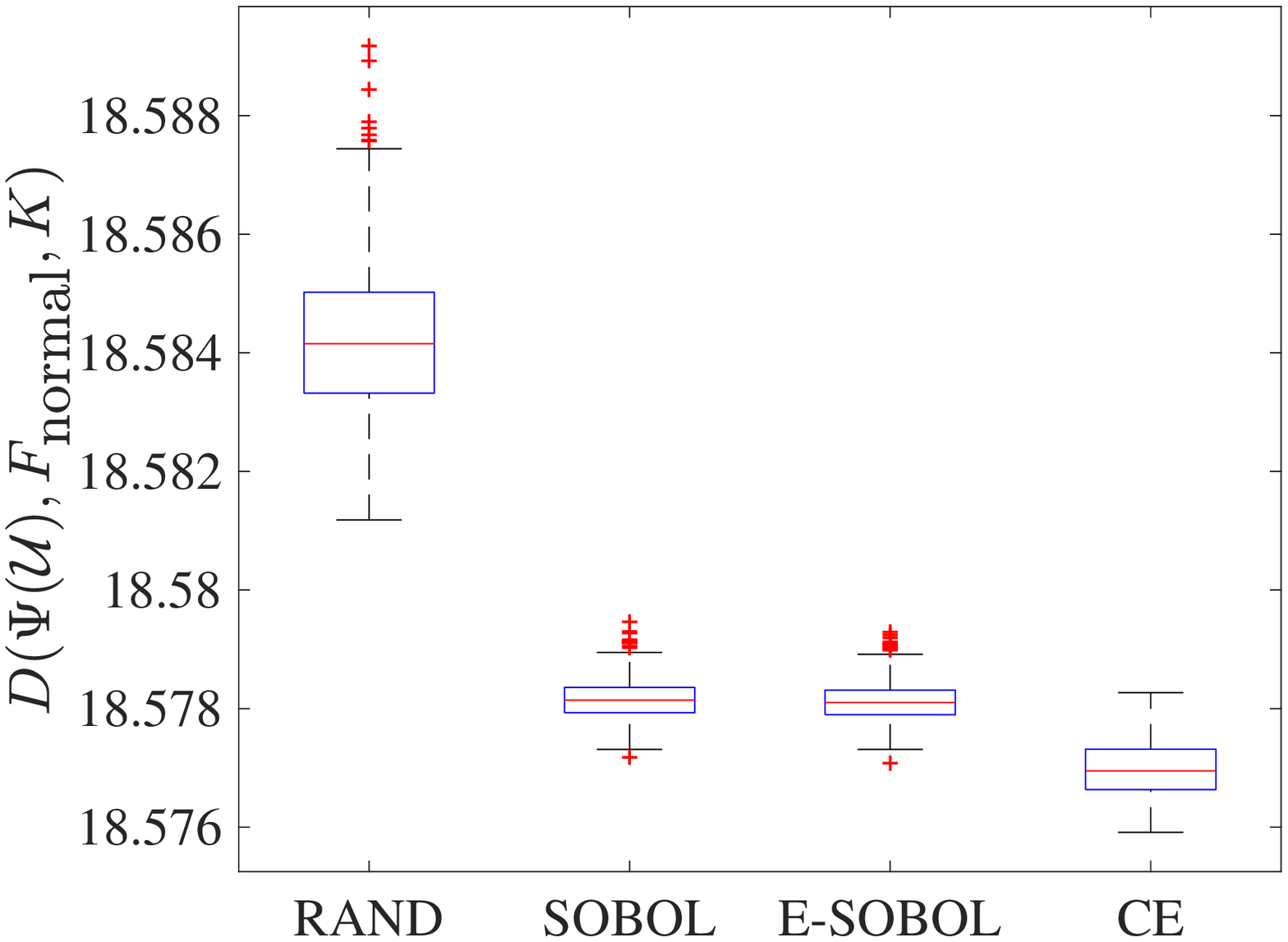}}}
\caption{Performance comparison of designs \label{fig:comparison}}
\end{figure}

Figure \ref{fig:SOBOLVsRAND} contains the boxplots of normal discrepancies corresponding to the four generators with $d=2$ and $N=32$. It shows that SOBOL, E-SOBOL, and CE all outperform RAND by a large margin. To better present the comparison between the better generators, in Figure \ref{fig:comparison} we generally exclude RAND.

We also report the average execution times for the four generators in Table \ref{tab:ExeTime}. All codes were run on a MacBook Pro with 2.4 GHz Intel Core i5 processor. The maximum number of iterations allowed is $M_{\max} = 200$.  Algorithm \ref{alg:coorex} converges within 20 iterations in all simulation examples.

 We summarize the results of our simulation as follows. 
\begin{enumerate}
\item
Overall, CE produces the smallest discrepancy.
\item 
When the design is relatively dense, i.e., $N/d$ is large, E-SOBOL and CE have similar performance.
\item
When the design is more sparse, i.e., $N/d$ is smaller, SOBOL and E-SOBOL have similar performance, but CE is superior to both of them in terms of the discrepancy. Not only in terms of the mean but also in terms of the \emph{range} for the $500$ designs generated.

\item CE requires the longest computational time to construct a design, but this is moderate.  When the cost of obtaining function values is substantial, then the cost of constructing the design may be insignificant.  
\end{enumerate}

\begin{table}[htbp]
  \centering
  \caption{Execution Time of Generators (in seconds)}
    \begin{tabular}{r*6{@{\ \ }c}}
    \toprule
         $d$ & $2$   & $3$   & $4$   & $6$   & $8$   & $10$ \\
  $N$ & $32$   & $64$   & $64$   & $128$   & $256$   & $512$ \\
    \hline
    RAND  & $3.22\text{E}{-5}$      &  $5.21\text{E}{-5}$     & $5.27\text{E}{-5}$      &  $9.92\text{E}{-5}$     & $2.48\text{E}{-4}$      & $ 5.32\text{E}{-4}$ \\
    SOBOL & $8.60\text{E}{-4}$      &  $0.10\text{E}{-2}$     &    $0.11\text{E}{-2}$   &  $0.16\text{E}{-2}$     & $0.21\text{E}{-2}$      &  $0.28\text{E}{-2}$\\
    E-SOBOL & $8.71\text{E}{-4}$  & $0.11\text{E}{-2}$      &     $0.12\text{E}{-2}$  &  $0.16\text{E}{-2}$     &   $0.23\text{E}{-2}$    & $0.32\text{E}{-2}$  \\
    CE    & $1.34\text{E}{-2}$      & $2.73\text{E}{-2}$      & $6.12\text{E}{-2}$      &   $0.24$    &   $1.04$    & $3.84$  \\
    \hline
    \end{tabular}%
  \label{tab:ExeTime}%
\end{table}%

\section{Discussion} \label{sec:discussion}

This chapter summarizes the three interpretations of the discrepancy. We show that for kernels and variable transformations satisfying conditions \eqref{eq:BigQYes}, variable transformations of low discrepancy uniform designs yield low discrepancy designs with respect to the target distribution.  However, for more practical choices of kernels, this correspondence may not hold.  The coordinate-exchange algorithm can improve the discrepancies of candidate designs that may be constructed by variable transformations. 

While discrepancies can be defined for arbitrary kernels, we believe that the choice of kernel can be important, especially for small sample sizes.  If the distribution has a symmetry, e.g. $\varrho(\vT(\vx)) = \varrho(\vx)$ for some probability preserving bijection $\vT:\Omega \to \Omega$, then we would like our discrepancy to remain unchanged under such a bijection, i.e., $D(\vT(\Xdes),\varrho,K) = D(\Xdes,\varrho,K)$.  This can typically be ensured by choosing kernels satisfying $K(\vT(\vt),\vT(\vx)) = K(\vt,\vx)$.   The kernel $K_\unif$ defined in \eqref{eq:L2kernel} satisfies this assumption for the standard uniform distribution and the transformation $\vT(\vx) = \vone - \vx$.  The kernel $K$ defined in \eqref{eq:OrigKernel} satisfies this assumption for the standard normal distribution and the transformation $\vT(\vx) = -\vx$.

For target distributions with independent marginals and kernels of product form as in \eqref{eq:prodkernel}, \emph{coordinate weights} \cite{DicEtal14a}*{Section 4} are used to determine which projections of the design, denoted by $\fraku \subseteq \{1, \ldots, d\}$, are more important.  The product form of the kernel given in \eqref{eq:prodkernel} can be generalized as 
\begin{equation*} 
     K_{\vgamma}(\vt,\vx)  = \prod_{j=1}^d\left[1+ \gamma_j \tK(t_j,x_j) \right].
\end{equation*}
Here, the positive coordinate weights are $\vgamma = (\gamma_1, \ldots, \gamma_d)$.
The squared discrepancy corresponding to this kernel may then be written as
\begin{align*}
      D^2(\Xdes,F,K_{\vgamma}) &= \sum_{\substack{\fraku \subseteq \{1, \ldots, d\} \\ \fraku \ne \emptyset}}\gamma_\fraku D^2_{\fraku} (\Xdes,\rho,K), \qquad \gamma_{\fraku} = \prod_{j \in \fraku} \gamma_j\\
    D^2_{\fraku} (\Xdes_\fraku,F_\fraku,K) & = c^{\abs{\fraku}} - \frac 2N \sum_{i=1}^N \prod_{j \in \fraku} h(x_{ij}) + \frac{1}{N^2} \sum_{i,k=1}^N \prod_{j \in \fraku} \tK(x_{ij},x_{kj}),
\end{align*}
where $c$ and $h$ are defined in \eqref{eq:chHdef}. Here, $\Xdes_\fraku$ denotes the projection of the design into the coordinates contained in $\fraku$, and $F_\fraku = \prod_{j \in \fraku} F_j$ is the $\fraku$-marginal distribution.  Each discrepancy piece, $D_{\fraku} (\Xdes_\fraku,F_\fraku,K)$, measures how well the projected design $\Xdes_\fraku$ matches $F_\fraku$. 

The values of the coordinate weights can be chosen to reflect the user's belief as to the importance of the design matching the target for various coordinate projections.  A large value of $\gamma_j$ relative to the other $\gamma_{j'}$ places more importance on the $D_{\fraku} (\Xdes_\fraku,F_\fraku,K)$ with $j \in \fraku$.  Thus, $\gamma_j$ is an indication of the importance of coordinate $j$ in the definition of $D(\Xdes,F,K_{\vgamma})$.  

If $\vgamma$ is one choice of coordinate weights and $\vgamma'=C\vgamma$ is another choice of coordinate weights where $C > 1$, then $\gamma'_\fraku = C^{\abs{\fraku}} \gamma_\fraku$.  Thus, $D(\Xdes,F,K_{\vgamma'})$ emphasizes the projections corresponding to the $\fraku$ with large $\abs{\fraku}$, i.e., the higher order effects. Likewise, $D(\Xdes,F,K_{\vgamma'})$ places relatively more emphasis lower order effects.  Again, the choice of coordinate weights reflects the user's belief as to the relative importance of the design matching the target distribution for lower order effects or higher order effects.

\bibliography{Ref.bib,FJH23,FJHown23}

\section*{Appendix}
We derive the formula in \eqref{eq:DisNormald>1} for the discrepancy with respect to the standard normal distribution, $\Phi$, using the kernel defined in \eqref{eq:OrigKernel}.  We first consider the case $d=1$. We integrate the kernel once:
\begin{align*}
\MoveEqLeft{\int_{-\infty}^\infty K(t,x) \, \dif \Phi(t)}\\
=&\int_{-\infty}^{\infty} \left(1+\frac{1}{2}|x|+\frac{1}{2}|t|-\frac{1}{2}|x-t|\right)\phi(t) \, \dif t\\
=& 1+ \frac{1}{\sqrt{2\pi}} + \frac{1}{2}|x|
-\frac{1}{2}\left[\int_{-\infty}^{x}(x - t)\phi (t) \, \dif t
+\int_{x}^{\infty} (t - x)\phi(t) \, \dif t\right]\\
=& 1+ \frac{1}{\sqrt{2\pi}} + \frac{1}{2}|x| - x [\Phi(x)-1/2] - \phi(x) .
\end{align*}
Then we integrate once more:
\begin{align*}
\MoveEqLeft{\int_{-\infty}^\infty \int_{-\infty}^\infty K(t,x) \, \dif \Phi(t) \dif \Phi(x)}\\
& =  \int_{-\infty}^{\infty} \left(1+ \frac{1}{\sqrt{2\pi}} + \frac{1}{2}|x| - x [\Phi(x)-1/2] - \phi(x) \right)\phi(x) \, \dif x\\
&= 1+ \sqrt{\frac{2}{\pi}} + \int_{-\infty}^{\infty} \{ - x \Phi(x)\phi(x)  +  [\phi(x)]^2 \} \, \dif x\\
&= 1+\sqrt{\frac{2}{\pi}}-\frac{1}{\sqrt{4\pi}}+\int_{-\infty}^{\infty}\frac{1}{2\pi}\mathrm{e}^{-x^2}\dif x =1+\sqrt{\frac{2}{\pi}}. 
\end{align*}

Generalizing this to the $d$-dimensional case yields 
\begin{gather*}
\int_{\reals^d\times \reals^d} K(\vx,\vt) \, \dif\Phi(\vx)\dif\Phi(\vt) = \left(1+\sqrt{\frac{2}{\pi}}\right)^d, \\
\int_{\reals^d}K(\vx,\vx_n) \, \dif\Phi(\vx) = \prod\limits_{j=1}^d \left[ 1+\frac{1}{\sqrt{2\pi}}+\frac{1}{2}|x_j|-x_j[\Phi(x_j)-1/2]-\phi(x_j)\right].
\end{gather*}
Thus, the discrepancy for the normal distribution is
\begin{eqnarray*}
&&D^2(\Xdes, \Phi, K)\\
&=& \left(1+\sqrt{\frac{2}{\pi}}\right)^d - \frac{2}{N}\sum\limits_{\vx\in P} \prod\limits_{j=1}^d\left[ 1+\frac{1}{\sqrt{2\pi}}+\frac{1}{2}|x_j|-x_j[\Phi_1(x_j)-1/2]-\phi(x_j)\right]\\
&&+\frac{1}{N^2}\sum_{\vx,\vt\in P}\prod_{j=1}^d \left[1+\frac{1}{2}|x_j|+\frac{1}{2}|t_j|-\frac{1}{2}|x_j-t_j|\right]. 
\end{eqnarray*}      

\end{document}